\documentclass[preprint,prd,nofootinbib]{revtex4}
\usepackage[dvips]{graphicx}
\usepackage{latexsym}
\long\def\ata#1{\footnote{E-mail: {\tt #1}}}

\newcommand{\La}{\mathcal{L}}				%Lagrangian
\newcommand{\A}{A_\mu}					%Ga	uge field
\newcommand{\PgP}{\bar\psi\gamma^\mu\psi}		%Fermion current
\newcommand{\tensorbilinear}{\bar\psi\frac{i}{2}(\gamma_\mu\buildrel\rightarrow\over\partial_\nu -
\gamma_\mu\buildrel\leftarrow\over\partial_\nu)\psi}		%Fermion bilinear with two indices
\long\def\slash#1{#1 \!\!\!/}						%Feynman slash

\newwrite\ffile\global\newcount\figno \global\figno=1

\def\writedef#1{}
\def\figin{\epsfcheck\figin}\def\figins{\epsfcheck\figins}

\def\epsfcheck{\ifx\epsfbox\UnDeFiNeD \message{(NO epsf.tex, FIGURES
WILL BE IGNORED)}
\gdef\figin##1{\vskip2in}\gdef\figins##1{\hskip.5in}% blank space
instead \else\message{(FIGURES WILL BE INCLUDED)}%
\gdef\figin##1{##1}\gdef\figins##1{##1}\fi} \def\figinsert{}
\def\ifig#1#2#3{\xdef#1{fig.~\the\figno} \writedef{#1\leftbracket
fig.\noexpand~\the\figno}%
\figinsert\figin{\centerline{#3}}\medskip\centerline{\vbox{\baselineskip12pt
\advance\hsize by -1truein\center\scriptsize{  Fig.~\the\figno.} #2}}
\bigskip\endinsert\global\advance\figno by1}
\def\endinsert{}
\begin{document}

\preprint{CALT-68-2521} \preprint{hep-th/0409189}

\title{~\\ Composite Mediators and Lorentz Violation \bigskip}
\author{Alejandro Jenkins\ata{jenkins@theory.caltech.edu}}
\affiliation{California Institute of Technology, Pasadena, CA 91125 \vspace{20 mm}}

\begin{abstract} \bigskip
We briefly review the history and current status of models of particle interactions in which massless mediators are given, not by fundamental gauge fields as in the Standard Model, but by composite degrees of freedom of fermionic systems.  Such models generally require the breaking of Lorentz invariance.  We describe schemes in which the photon and the graviton emerge as Goldstone bosons from the breaking of Lorentz invariance, as well as generalizations of the quantum Hall effect in which composite excitations yield massless particles of all integer spins.  While these schemes are of limited interest for the photon (spin 1), in the case of the graviton (spin 2) they offer a possible solution to the long-standing UV problem in quantum linear gravity. \vspace{15 mm}

\begin{center}
{\it Contributed to the Third Meeting on \\
CPT and Lorentz Symmetry \\
University of Indiana at Bloomington \\
August 4-7, 2004} \\
\end{center}

\end{abstract}

\maketitle

\section{Why local gauge invariance?}

The Dirac Lagrangian for a free fermions, $\La = \bar\psi (i \slash\partial - m) \psi$ is invariant under the global $U(1)$ gauge transformation $\psi \mapsto \exp({i \alpha})\psi$.  In the established model of quantum electrodynamics, this Lagrangian is transformed into an interacting theory by making the gauge symmetry local: the phase $\alpha$ is allowed to be a function of the space-time point $x^\mu$.  This requires the introduction of a gauge field $\A$ with the the transformation property $\A \mapsto \A + \partial_\mu \alpha$, and the use of a ``covariant derivative'' $D_\mu = \partial_\mu - i \A$ instead of the usual derivative $\partial_\mu$.  The generalization to non-abelian gauge groups is well known, as is the Higgs mechanism to spontaneously break the gauge invariance and give the field $\A$ a mass.

A deeper insight into the physical meaning of local gauge invariance comes from realizing that a massless spin 1 particle, having no rest frame, cannot have its spin point along any axis other than that of its motion.  Therefore, it has only two polarizations.  By describing it as Lorentz vector $\A$ (which has three polarizations) a mathematical redundacy is introduced.  This redundancy is local gauge invariance.  Something like it must appear in any Lorentz invariant theory of a massless spin 1 field coupled to matter.  In general relativity, the graviton is a massless particle with two polarizations, but it is described by a spin 2 field, which would ordinarily have five polarizations.  This redundancy leads to diffeomorphism invariance, a symmetry analogous to local gauge invariance in the spin 1 case.  (See, for instance, chapter 5.9 in \cite{Weinberg} and chapter III.3 in \cite{Zee}.)

Modern particle theory is based on local gauge invariance, and it has been shown that gauge theories have the very attractive feature that they are always renormalizable \cite{tHooft}.  But there is no clearly compelling {\it a priori} reason to impose local gauge invariance as an axiom.  Also, it might appear unsatisfactory that our mathematical description of physical reality should be inherently redundant: local gauge invariance, unlike a true physical symmetry, does not mean that different physical configurations have the same behavior.  Rather, it means that different field configurations represent exactly the same physics \cite{Zee}.  Finally, local gauge invariance as a guarantee of renormalizability works only for spin 1.  It is well known that quantizing $h_{\mu\nu}$ in linear gravity does not produce a perturbatively renormalizable field theory.

\section{Goldstone photons}

Before quantum chromodynamics (QCD), an $SU(3)$ gauge theory, was accepted as a model for the strong nuclear force, Nambu and Jona-Lasinio (NJL) proposed a scheme in which protons and neutrons in nuclei would interact strongly by exchanging composite massless particles associated with the spontaneous breaking of the chiral symmetry $\psi \mapsto \exp{(i\alpha\gamma^5)} \psi$ \cite{NJL}.  That is, in their model, the pions were composite Goldstone bosons in a theory whose only fundamental fields were fermions.

Shortly after the NJL model was published, Bjorken proposed using a similar idea to account for QED without postulating $U(1)$ local gauge invariance \cite{Bjorken1,Bjorken2}.  He suggested that a theory with only self-interacting fermions might spontaneously break Lorentz invariance, yielding composite Goldstone bosons that could act as the mediators of the electromagnetic force.

Conceptually, a useful way of understanding Bjorken's proposal is to think of it as as a resurrection of the lumineferous \ae ther \cite{Nambu1,Nambu2}:  ``empty'' space is no longer really empty.\footnote{Taylor and Wheeler declare in \cite{Wheeler} that one can think of Einstein's special relativity (and therefore Lorentz invariance) simply as the statement that empty space is {\it really} empty.}  Instead, the theory has a non-vanishing vacuum expectation value (VEV) for the current $j^\mu=\PgP$.  This VEV, in turn, leads to a massive background gauge field $\A \propto j_\mu$, as in the well-known London equations for the theory of superconductors.\footnote{In Bjorken's work, $\A$ is just an auxiliary or interpolating field.  Dirac had discussed somewhat similar ideas in an earlier paper \cite{Dirac}, but, amusingly, he was trying to write a theory of electromagnetism with only a gauge field and no fundamental electrons.  In both the work of Bjorken and the work of Dirac, the proportionality between $\A$ and $j_\mu$ is crucial.}  Such a background spontaneously breaks Lorentz invariance and produces three massless excitations of $\A$ (the Goldstone bosons) proportional to the changes $\delta j_\mu$ associated with the three broken Lorentz transformations.

Two of these Goldstone bosons can be interpreted as the usual transverse photons.  The meaning of the third photon remains problematic.  Bjorken originally interpreted it as the longitudinal photon in the temporal-gauge QED, which becomes identified with the Coulomb force (see also \cite{Nambu1}).  More recently, Kraus and Tomboulis have argued that the extra photon has an exotic dispersion relation and that its coupling to matter should be suppressed \cite{KrausTomboulis}.

\section{Goldstone gravitons}

The problem of the nonrenormalizability of linear gravity in the usual quantum field theory has been one of the major motivations for research in string theory, quantum loop gravity and other proposed theories of quantum gravity currently at the forefront of fundamental theoretical particle physics.  An early suggestion for solving the problem of linear gravity in the UV was to make the graviton a composite degree of freedom of the low-energy regime.  Weinberg and Witten, however, put an end to much of the speculation in this direction by using a strikingly simple argument to show that Lorentz invariant field theories with a Lorentz covariant energy-momentum tensor $T_{\mu\nu}$ do not admit massless degrees of freedom, either fundamental or composite, with spin greater than 1 \cite{WeinbergWitten}.  (GR has a massless spin 2 particle, the graviton, because the energy-momentum of the gravitational field is given by a non-covariant $T_{\mu\nu}$.  Local conservation of the stress-energy tensor for linear gravity prevents the graviton field $h_{\mu\nu}$ from transforming like a Lorentz tensor.)

It is, however, possible to invoke a mechanism similar to the one described by Bjorken in order to obtain composite gravitons as the Goldstone bosons of spontaneously broken Lorentz invariance.  If a field $h_{\mu\nu}$ acquires a non-zero VEV, then the $SO(3,1)$ Lorentz symmetry would be broken to nothing, generating six Goldstone bosons.  The VEV of $h_{\mu\nu}$ can be thought of as proportional to some non-vanishing tensor bilinear in the background, such as $\tensorbilinear$ \cite{KrausTomboulis}.  The question of how to obtain such a VEV remains problematic \cite{Jenkins}.

Of the five massless excitations from the breaking of Lorentz invariance, two could be identified with the helicities of the graviton, while the other four should presumably have their interactions with matter suppressed.

\section{Generalization of the quantum Hall effect}

In 1983, Laughlin explained the observed fractional quantum Hall effect in two-dimensional electronic systems by showing how such a system could form an incompressible quantum fluid whose excitations have charge $e/3$ \cite{Laughlin}.  That is, the low-energy theory of the interacting electrons in two spatial dimensions has composite degrees of freedom whose charge is a fraction of that of the electrons themselves.  In 2001, Zhang and Hu used techniques similar to Laughlin's to study the composite excitations of a higher-dimensional system \cite{ZhangHu}.  They imagined a four-dimensional sphere in space, filled with fermions that interact via an $SU(2)$ gauge field.  In the limit where the dimensionality of the representation of $SU(2)$ is taken to be very large, such a theory exhibits composite massless excitations of integer spin 1, 2 and higher.

Like other theories from solid state physics, Zhang and Hu's proposal falls outside the scope of Weinberg and Witten's theorem because the proposed theory is not Lorentz invariant:  the vacuum of the theory is not empty and has a preferred rest-frame (the rest frame of the fermions).  However, the authors argued that in the three-dimensional boundary of the four-dimensional sphere, a relativistic dispersion relation will hold.  One might then imagine that the relativistic, three-dimensional world we inhabit might be the edge of a four-dimensional sphere filled with fermions.  Photons and gravitons would be composite low-energy degrees of freedom, and the problems currently associated with gravity in the UV would be avoided.  The authors also argue that massless bosons with spin 3 and higher might naturally decouple from other matter, thus explaining why they are not observed in nature.

\section{Outlook}

The model of the strong interactions proposed by Nambu and Jona-Lasinio was eventually superseded by QCD, and Bjorken's proposal for composite photons was similarly overtaken by the steady rise of local gauge invariance as a sacred principle of theoretical particle physics.  But NJL survived, in modified form, as the basis of chiral perturbation theory, and it is possible that Bjorken's model might make a comeback, in the way that the old Kaluza-Klein model was resurrected by modern theories with extra dimensions, such as string theory.  A theory with composite gravitons currently seems especially appealing, given the interest in addressing the UV problems of gravity in the the quantum field theory context.  Much work remains to be done in this area.

One interesting issue is whether models of composite massless mediators might be associated with observable violations of Lorentz invariance.  References \cite{Dirac}, \cite{Bjorken1}, and \cite{Nambu1} claimed that the proposed Lorentz violation was purely formal and had no observable consequences, since it appeared only as a VEV of $\A$, which could be gauged away.  However, it has recently become apparent that in many such models the Lorentz violation is physical \cite{Bjorken2,KrausTomboulis,Nambu2,Jenkins} and associated with the presence of an ``\ae ther'' given by a non-empty Dirac sea of fermions in the background that also introduces a chemical potential \cite{Nambu2,Jenkins}.  On the other hand, the model proposed in \cite{ZhangHu} rescues special relativity in the boundary region in which they imagine our universe is located.

\section*{Acknowledgments}

The author thanks J.D.\ Bjorken and Y.\ Nambu for fruitful exchanges.  This work was supported in part by an R.A.\ Millikan graduate fellowship from the California Institute of Technology.

\end{document}